\documentstyle[sprocl]{article}

\def\Journal#1#2#3#4{{#1} {\bf #2}, #3 (#4)}

\def\PRD{{\em Phys. Rev.} D}

\def\CQG{{\em Class. Quant. Gravity} }

\def\ra{\rightarrow}

\def\be{\begin{equation}}
\def\ee{\end{equation}}
\def\bea{\begin{eqnarray}}
\def\eea{\end{eqnarray}}

\begin{document}

\title{
   CHAOTIC  QUANTIZATION OF CLASSICAL GAUGE FIELDS 
}

\author{T.S. Bir\'o,}

\address{MTA KFKI Res.Inst. for Part. and Nucl. Phys. \\
H-1525 Budapest, Pf. 49,  Hungary \\E-mail: tsbiro@sunserv.kfki.hu} 

%

\author{S. G. Matinyan and B. M\"uller}

\address{Department of Physics, Duke University \\
Durham, NC 27708-0305, USA\\E-mail: muller@phy.duke.edu}


\maketitle\abstracts{ 
 We argue that higher dimensional classical, nonabelian
 gauge theory may lead to a lower dimensional quantum
 field theory due to its inherent chaotic dynamics which
 acts like stohastic quantization. 
 The dimensional reduction is based upon magnetic
 screening effects analogous to that in non-abelian plasmas.
 After reviewing properties
 of classical chaos in numerical investigations of lattice gauge  
 theory we discuss this mechanism in details.
}

Since the inclusion of gravity into quantum field theory 
by quantizing the gravitational interaction is still unsolved,
the speculative counterpart of this unification attempt,
namely basing quantum field theory onto classical grounds,
occurs. Such an alternative has been suggested by t'Hooft\cite{Hoo},
and has been discussed in some recent considerations
exploring the correspondence between 
gravity in $D+1$ dimensions AdS space and
conformal field theory (CFT). There has been, however,
no physical mechanism outlined yet, how the
dimensional reduction (others than an unexplained compactification
in spacelike dimensions) manifests itself in nature.

In this lecture we present such a mechanism relying
on a {\em classical}, Yang-Mills type theory in $D+2$
dimensions, which we learned to be chaotic. The chaotic
dynamics leads to self-ergodization, equipartioning the
energy and thus leading to a thermal state: this is the
energy -- temperature correspondence ($E \rightarrow T$).
The  dynamics of selected long wavelength modes can effectively
be described by a Langevin equation including thermal
noise: by this view the study of a given field configuration
is generalized to the investigation of a distribution
over possible (and realized) field configurations.
The dynamics of this distribution
-- followed by the chaotic system in a rapid sequence
of configurations visited by a generic solution curve 
in one of the extra time dimensions over $D$ space dimensions --
is described by a (still $D+2$-dimensional) Fokker-Planck
equation. This change of view can be noted for short as
$A \rightarrow P[A]$. Finally the stationary solution
of this Fokker-Planck equation, which would occur as physical
reality for slow observers, represents the ground state
of a $D+1$-dimensional quantum field theory belonging to
the magnetic sector of the original Yang-Mills action,
expressed in the lower dimension only. This relation is
analogous to that of a stohastic quantization, but here
the thermal bath is not given from outside, it is an
intrinsic part of the higher dimensional, classical
dynamics. This step allows for interpreting Planck's
constant in the lower dimensional world: 
$H_{D+2}/T \rightarrow S_{D+1}/\hbar$.

The outline of this lecture is as follows: first some
questions relevant to basic concepts of space, time and
quantization are listed in order to motivate our search
for the above described scenario of natural ``self-quantization''.
Then  results on lattice and continuum Yang-Mills theory
are reviewed, which were carried out in past years investigating
the chaotic nature of the classical dynamics of such
systems\cite{BMM}. In particular extrapolations to the continuum limit
are taken care of. After this more or less extended review
we present our considerations to the quantization mechanism
due to higher dimensional chaos relating the $D+1$-dimensional
Planck's constant to a $D+2$-dimensional temperature
and a fiducial length-scale. A discussion of space dimensionalities
and internal gauge degrees of freedom necessary for chaotic
dynamics closes this part. Finally some speculations
are presented, whether and how quantum mechanics can ever be
experimentally falsified and if, at which energy scale.

\section{Questions}

Considering Quantum Field Theory (QFT) as a lower
dimensional {\em boundary} of a higher dimensional classical
field theory (ClassFT) assumes spacelike higher dimensions.
A chaotic quantization assumes that higher dimensions
(or at least one of them) are timelike. Here we list
questions which motivated our considerations of a
chaotic quantization mechanism or led from these 
considerations to relations to other, partially long known,
but not yet satisfactorily answered questions of
fundamental quantum mechanics.

{\em Is (can be) something behind QFT?}
If yes at some instance there should be experiments 
which cannot be expalined by the principles of quantum mechanics
(such as interference and linearity).

{\em Is (can be) stohastic quantization natural?}
If yes, this brings us close to a many world interpretation
of quantum mechanics, with the only difference
that these copies of the universe would not be
parallel but rather follow a sequence (``second'' time
coordinate).

{\em Is (can be) given a timescale, beyond which quantum
mechanics is no more valid?}
If yes, for very short times (at very high energy)
Bell's inequality would be satisfied again corresponding
to classicality in higher dimensions.

{\em Is (can be) higher dimension a hidden parameter?}
If yes, such higher dimensional theories are beyond
quantum theory and Planck's constant in the
phenomenological world becomes derivable from
(chracteristically more than one) parameters of
the underlying theory.

{\em Is (can be) time two dimensional?}
If yes, if there is an extra timelike dimension,
time is no more one-dimensional in the underlying
theory and the possibility of ordering (past - future)
is lost. Physical time will be given birth first by
quantization.

{\em Is (can be) chaotic dynamics a self-driven
mechanism for quantization?}
If it is so, the observed world of quantum field
theories is experienced only in the infrared limit
of the underlying theory; making only ergodically
distributed average information (wave function)
available for the ``slow'' experimentators.

{\em Is (can be) spacetime a lattice?}
This is of course unlikely, and cannot be expected
before the Planck scale. Even if it were so, then
we would not have any reason to give preference
to a cubic lattice (or to a hexagonal or to any other
one). The continuum (ultarviolet) limit of classical lattice
theories is therefore particularly important.

{\em Is (can be) any classical theory insane enough to
lead to quantum mechanics? }
This question is due to H.~B.~Nielsen.
It implies that the answer is better no. Inspite this warning,
however, we can investigate, whether the
classical dynamics of (certain) classical field
theories possess at all characteristica which make
them fit for the role of stohastic quantizator.
In the next section we therefore review basic results
of the study of classical chaos in lattice gauge theory 
and its continuum limit carried out in past years,
in particular from the viewpoint of a possible
applicability to chaotic quantization.

\section{Classical Chaos in Lattice Gauge Theory}

First studies of the chaotic dynamics of non-abelian
gauge theories occured in the 1980-s, they were
restricted to a few (infrared) degrees of freedom.
Even the simplest Hamiltonian,
\be
H = \frac{1}{2} \left(\dot{A}_1^2\right)^2
+ \frac{1}{2} \left(\dot{A}_2^1\right)^2
+ \frac{1}{2} g^2 \left( A_1^2 A_2^1 \right)^2
\ee
led to chaotic motion. (Here $A_1^2$ and $A_2^1$ are
SU(2) vector potential components).
The classical solutions are scaling and everywhere chaotic.
The surface of constant potential energy is hyperbolic, 
so the turning points of classical trajectories 
constitute a hypersurface with overall negative curvature.
This system is part of the SU(2) gauge theory, it
resides in the infrared sector.

In the 1990-s numerical studies on lattice gauge
systems with U(1), SU(2) and SU(3) gauge groups has been
carried out, up to linear size of $N=50$. Here the
classical dynamics is governed by a lattice Hamiltonian,
written in a per link version as
\be
H =  \sum \left[\frac{1}{2} \langle P, P \rangle 
+ \left( 1 - \frac{1}{4} \langle U, V \rangle \right) \right],
\ee
with the general scalar product notation
$$ \langle A, B \rangle = \frac{1}{2} {\rm tr} (AB^{\dag}) $$
for SU(2) matrices $A$ and $B$, $U$ being the adjoint
group variable describing a compact version of the
vectorpotential on each lattice link,
\be
U = \exp ( ig a^{\mu} A_{\mu}^a T^a ),
\ee
and $V$ the complement link variable constructed from products
of $U$ variables on links closing an elementary plaquette
with the selected link. Besides the Hamiltonian governing
the dynamics, the unitarity ($\langle U, U \rangle = 1$)
and orthogonality ($\langle P, U \rangle = 0$) is taken
care of. There is a non-trivial conserved quantity
expressing Gauss' law as well:
\be
\Gamma = \sum_+ P U^{\dag} - \sum_- U^{\dag} P,
\ee
defined on the joints of links. The $+$ sum belongs to
links starting at, while the $-$ sum to links ending at
a given joint. With update algorithms specifically designed
to satisfy conservation laws the Hamiltonian dynamics
of such lattice systems can be followed. Chaos reveals
itself in divergence properties of nearby configurations.

In order to give words ``nearby'' and ``diverging'' a meaning
one has to use a distance measure in the phase space of
lattice gauge field configurations. The simplest gauge
invariant choice is given by a sum of local deviations
in the magnetic energy part:
\be
d = \frac{1}{N_{{\rm link}}} \sum_{{\rm link}}
\left| \langle U, V(U) \rangle - \langle U', V(U') \rangle \right|.
\ee
Starting with an adjacent pair of configurations $U$ and $U'$
the initial distance $d_0$ let be small. The Lyapunov
exponent measures the long term divergence of such configurations:
\be
\lambda = \lim_{t\ra \infty} \lim_{d_0 \ra 0}
\frac{1}{t} {\rm ln} \frac{d(t)}{d_0}.
\ee
Since the lattice Hamiltonian is conservative, the sum of
all Lyapunov exponents is zero, but the sum of positive
ones gives a good measure of the entropy generation:
\be
h_{KS} = \sum_{\lambda > 0} \lambda_i.
\ee
As entropy is generated such systems ergodize without an
external heat bath; the initially given energy becomes
equipartioned in a chracteristic time. Assuming an ideal
thermal state the correspondence between energy per degree
of freedom $E$ and temperature $T$ can be checked, as well as
the maximum entropy $S$. This way a characteristic 
thermalization time is given by $\tau = S/h_{KS}.$

The numerical studies of the chaos
in lattice gauge systems gave the following main results:
\begin{itemize}

\item	{\em Lyapunov exponents:} the total spectrum of Lyapunov
	exponents for smaller systems, and a correspondence
	between the maximal Lyapunov and the energy per degree
	of freedom in general were obtained, \hbox{$ \lambda_0 \propto  g^2E$}
	for all non-abelian gauge groups, valid also  
        in the continuum ($a \ra 0$) limit.

\item	{\em Scaling:} these result scale with system size, giving
	a numerical allusion to the extensivity of the
	chaos phenomenon. It relies on the infrared sector.

\item	{\em Ergodicity:} the ergodizing speed can be estimated by
	measuring the Kolmogorov entropy, its scaling
	properties with system size and observation time
	revealed that the ensemble of chaotic evolutions
	looks like motion in a Gaussian noise.

\item	Uniformly random configurations and those selected
	by quantum Monte Carlo programs has been compared.
	From the viewpoint of chaotic behavior they are found to
	be interchangable. The strength of chaoticity 
	nevertheless correlates with the confining phase.

\end{itemize}

Beyond and parallel to the numerical studies, analytical
results has also been obtained in high temperature field theory
(resummation, study of transport properties). The most
relevant results from this extended research for the investigation
of chaotic quantization are the followings:

\begin{itemize}
\item	{\em Length scale hierarchy:} a hierarchy of thermal
	($\hbar/T$), electric ($\sqrt{\hbar}/gT$) and
	magnetic ($1/g^2T$) screening lengthes is established.

\item	{\em Dimensional reduction:} at very high temperature 
	the effective
	theory resembles a field theory in a lower dimension,
	with a dimensionally transmutated effective coupling.

\item	{\em Noise:} an effective Langevin equation describing
	evolution in noisy background medium has been derived
	for the infrared plasma modes, valid for times
	$t \gg 1/g^2T$ by B\"odeker\cite{Boe}.

\end{itemize}

These cited results are the necessary ingredients to combine
a mechanism for chaotic quantization. It is presented in the
next section.

\section{The Mechanism of Chaotic Quantization}

{\bf Step 1.} The first step of the chaotic quantization
mechanism is the connection between a $D+2$-dimensional
Yang-Mills theory and the corresponding Langevin equation:
$$ YM_{D+2} \longrightarrow {\rm Langevin}_{D+2}. $$
This accounts for the equipartition of energy
($ E \ra T$) replacing the study of the chaotic
Hamiltonian,
\be
{\cal H} = \frac{1}{2} \dot{A}^2 + \frac{1}{2}
(D \times A)^2,
\ee
by solving the Langevin equation
\be
\sigma \dot{A} = - D \times B + \eta,
\ee
with $B = D \times A$ magnetic field and $\eta$ white
noise corrrelated as
\be
\langle \eta \eta' \rangle = 2\sigma T \delta().
\ee
This resummed, finite-T perturbation theory result occuring
first at the two-loop level, has been recently published
\cite{ASY}. It is interesting to note, that the
formal usage of (chromo)electric field by
\be
\dot{E} =  D \times B
\ee
(together with $E = \dot{A}$ ) renders this Langevin
equation to one describing a simple Brownian motion:
\be
\dot{E} + \sigma E = \eta.
\ee
It is easy to derive and has been since long known that Brownian
motion leads to equipartition of the energy,
\be
{\cal H}_{\em elec} = \frac{1}{2} E^2,
\ee
in a characteristic time $\tau = 1/(2\sigma)$:
\be
{\cal H}_{\rm elec} \ra  \frac{1}{2} T.
\ee

Since the effective Langevin equation refers to a temperature
and to the coefficient $\sigma$, called color conductivity,
which depends on the ratio of magnetic and electric
screening lengthes, this relation can be generalized.
First of all, all these length scales and therefore the
whole Langevin equation approach {\bf works as well
in classical field theory}.

In the following table we list the corresponding quantities
in the quantum field theory and in the classical lattice
gauge theory case, respectively. 
We consider in order the magnetic and electric screening
lengthes, the squared plasma frequency and leading order
damping, the color conductivity and the hierarchy condition.
The paralells are intriguing.


\begin{table}[h]
\caption{ Comparison of high-T QFT and classical LGT quantities
\label{tab:class-qm}}
\vspace{0.2cm}
\begin{center}
\footnotesize
\begin{tabular}{l|c|c}
\hline
{} & QFT & Class. LGT
\\
\hline
 $d_{{\rm mag}}$ &  $1/(g^2T)$ &  $1/(g^2T)$ \\
\hline
 $d_{{\rm el}}$  &  $\sqrt{\hbar}/(gT)$ & $\sqrt{a/(g^2T)}$ \\
\hline
 $\omega_p^2$  &  $g^2T^2/{\hbar}$ & $g^2T/a$ \\
\hline
 $\gamma_p$  &  $g^2T$ &  $g^2T$ \\
\hline
 $\sigma$  &  $\frac{T/\hbar}{{\rm ln} (1/g^2\hbar) }$ &
   $\frac{1/a}{{\rm ln} (1/g^2Ta) }$ \\
\hline
 $d_{{\rm mag}} \gg d_{{\rm el}}$  &  $g^2\hbar \ll 1$ &
   $a \ll 1/(g^2T)$ \\
\hline
\end{tabular}
\end{center}
\end{table}


All these results are equivalent if one considers the
following conjecture:
$$ \hbar = a T. $$

The characteristic self-ergodization time is in
the same order of magnitude either expressed due to
the color conductivity or due to the Lyapunov exponent
\be
\tau = \frac{1}{2\sigma} \approx a \, {\rm ln}  \frac{1}{g^2Ta}
\approx a \, {\rm ln} \frac{1}{a\lambda_0}.
\ee

\vspace{50mm}
{\bf Step 2.} The solution of the Langevin equation is an
ensemble of solutions, since the noise is known only
statistically. This ensemble can be described equivalently
by the solution of a corresponding Fokker-Planck
equation:
$$ {\rm Langevin}_{D+2} \ra {\rm Fokker-Planck}_{D+2}.$$
Instead of equation(s) for the field confioguration $A$
this equation is for the distribution of solution
field configurations $P[A]$:
\be
\sigma \frac{\partial}{\partial x_{D+2}} P[A]
= \int \! d^{D+1}x \, 
\frac{\delta}{\delta A} \left( T \frac{\delta P}{\delta A}
 + \frac{\delta W}{\delta A} P \right).
\ee
This is an integro - differential variational equation, analog to
the one used in stohastic quantization. The driving
term is the deterministic force term of the Langevin
equation,
\be
\frac{\delta W}{\delta A} = - D \times B,
\ee
and hence
\be
W[A] = - \int \! d^{D+1}x \, \frac{1}{4} F_{ij}F^{ij}.
\ee
This ``potential'' energy is a $D+1$ dimensional integral
of the magnetic energy density alone, but can be related
to a $D+1$-dimensional Yang-Mills action. This is done in
the next step.

\vspace{5mm}
{\bf Step 3.} 
The stationary solution of the Fokker-Planck equation is
what a slow observer experiences: the effective theory
at large times features this distribution.
$$ {\rm Fokker-Planck}_{D+2} \ra 
{\rm Euclidean \, Path \, Integral}_{D+1}.$$
The $t \ra \infty$ solution is simply a Boltzmann-weight
\be
P[A] \ra e^{-W[A]/T}.
\ee
Interpreting this as proper weights of an euclidean field theory  
with the $D+1$-dimensional action, we identify
\be
\frac{W[A]}{T} = \frac{S_{D+1}}{\hbar_{D+1}},
\label{BOLTZMANN-FEYNMAN}
\ee
and
\be
S_{D+1} = - \frac{1}{4} \int \! dx_{D+1} \int \! d^Dx \,
\hat{F}^{ik}\hat{F}_{ik}.
\ee
Here $\hat{F}_{ik}$ stands for the field strength tensor
of the $D+1$-dimensional Yang-Mills action, consisting of
the integral of the magnetic part of the original, classical
Yang-Mills action, up to a scaling. This scaling is related
to the reduction of dimensionality of the integrals,
\be
\int \! d^{D+2}x \, \ldots = \, a \int \! d^{D+1}x \, \ldots
\ee
and requires a re-definition of the field strength
\be
F_{D+2}^{ik} = \sqrt{a} \hat{F}_{D+1}^{ik}.
\ee
Finally, comparing this with eq.(\ref{BOLTZMANN-FEYNMAN})
we arrive at
\be
\hbar_{D+1} = a \, T_{D+2}.
\ee
The Planck constant is expressed by parameters of the higher
dimensional theory. This result is consequent with the
comparison of lattice and continuum Yang-Mills screening properties
listed in Table 1, and therefore gives a good reason to
accept that it is valid also in the continuum limit.

\section{Speculations}

Finally we birefly mention a few speculations about further
possible extensions of this study. First the question
arises: is there any guiding principle to decide
whether a classical field theory is chaotic? A simple
consideration shows, that a counting of space-time dimensions
and gauge group degrees of freedom can give at least a
minimum estimate for this.

We count the unconstrained degrees of freedom in phase
space in the infrared ($k=0$ mode) limit. For a $D+2$
dimensional theory with ${\cal N}$ internal gauge bosons,
the naive phase space is $2(D+1){\cal N}$ dimensional
according to the variables $A_i^a(t)$ and $E_i^a(t)$.
Gauss law and the rest gauge symmetry in the Hamiltonian
approach restricts $2{\cal N}$ of them, the $SO(D,1)$
rotational (Lorentz-) symmetry -- active both for angles
and angular momenta -- further $2 D(D+1)/2$. The rest
dimensionality of the phase space becomes
\be
n = 2(D+1){\cal N} - 2{\cal N} - 2 \frac{D(D+1)}{2}
= D ( 2 {\cal N} - D - 1).
\ee
Chaotic dynamics of long wavelength modes can be expected
if $n \ge 3$. (Actually $n$ is even for conservative
systems.) This way the ${\cal N}=1$ case, e.g. U(1),
is not chaotic for any space dimension ($n = D(1-D)$).
${\cal N}=3$, e.g. SU(2), is chaotic for space dimensions
between 1 and 4:
\be
n = D(5-D) \ge 3, \qquad {\rm for} \qquad 1 \le D \le 4.
\ee
For gravitational theories, with ${\cal N}$ counted
from the higher dimensional Poincare group,
\be 
{\cal N} = \frac{(D+1)(D+2)}{2} 
\ee
we arrive at
\be
n = D(D+1)^2.
\ee
It can be chaotic already for one space dimension $D \ge 1$
(3 dimensional classical theory). 

Finally we note that considering two-dimensional time as
primordial also has been occured in other context:
I. Bars arrives at the one-dimensional (phenomenological)
time by gauge fixing in the higher dimensional (M)
theory\cite{Bar}. This has consequences for energy spectra and
dispersion relations in the lower dimensional theory and
serves to explore symmetries which might be non-trivial
(alike $O(4,2)$ in special relativity).

In the chaotic quantization mechanism time is also two
dimensional to begin with. Along one direction a ``physical''
time arises after quantization -- correspondingly the
fast evolution over chaotic field configuration
sequences must occur in higher time directions
(at least locally) orthogonal to this time curve.
The characteristic time in this orthogonal direction
is given by the inverse color conductivity.

Summerizing we proposed the chaotic quantization mechanism
underlying quantum field theory. This mechanism relies on
a time and length scale, which can be much larger, than
the Planck scale: $1/(2\sigma) \gg a$. Some consequences
for the interpretation of quantum mechanics also has been
mentioned.


\section*{Acknowledgments}
Discussions with H.B. Nielsen are gratefully acknowledged.
This work has been supported by the American Hungarian Joint
Fund JFNr. 649 and the Hungarian National Research Fund
OTKA T029158.




\section*{References}
\begin{thebibliography}{99a}

\bibitem{Hoo}G. t'Hooft, \Journal{\CQG}{16}{3263}{1999}.

\bibitem{BMM}C T. S. Bir\'o, B. M\"uller, S. G. Matinyan:
{\em Chaos and Gauge Field Theory,}
(World Scientific, Singapore, 1995).

\bibitem{Bar}I. Bars, \Journal{\PRD}{62}{046007}{2000}.


\bibitem{Boe}
D. B\"odeker,
{\em Phys. Lett. B}{\bf 426}, 351 (1998);

\bibitem{ASY}
P. Arnold, D.T. Son, and L.G. Yaffe,
{\em Phys. Rev. D}{\bf 59}, 105020 (1999);
{\bf 60}, 025007 (1999).



\end{thebibliography}
\end{document}